\documentclass[graybox,natbib,nosecnum]{svmult}
\bibpunct{(}{)}{;}{a}{}{,} 

\pdfoutput=1   

\usepackage{mathptmx}       
\usepackage{helvet}         
\usepackage{courier}        
\usepackage{type1cm}        

\usepackage{makeidx}         
\usepackage{graphicx}        
\usepackage{multicol}        
\usepackage[bottom]{footmisc}
\usepackage[normalem]{ulem}	
\usepackage{hyperref}  
\usepackage{xspace}
\usepackage{xcolor}
\usepackage{soul}   
\DeclareFontFamily{U}{mathb}{\hyphenchar\font45}
\DeclareFontShape{U}{mathb}{m}{n}{
      <5> <6> <7> <8> <9> <10> gen * mathb
      <10.95> mathb10 <12> <14.4> <17.28> <20.74> <24.88> mathb12
      }{}
\DeclareSymbolFont{mathb}{U}{mathb}{m}{n}
\DeclareMathSymbol{\Earth}{3}{mathb}{"43}

\newcommand{\RE}{R$_{\rm \Earth}$\xspace}
\newcommand{\ME}{M$_{\rm \Earth}$\xspace}

\newcommand\rev[1]{\textcolor{black}{#1}} 

\makeindex             

\begin{document}

\title*{Assessing the Interior Structure of Terrestrial Exoplanets with Implications for Habitability}
\titlerunning{Terrestrial Exoplanet Interiors} 
\author{Caroline Dorn, Dan J. Bower, and Antoine Rozel}
\institute{Caroline Dorn \at University of Z\"{u}rich, \email{cdorn@physik.uzh.ch}
\and Dan J. Bower \at University of Bern, \email{daniel.bower@csh.unibe.ch}
\and Antoine Rozel \at ETH Z\"{u}rich, \email{antoine.rozel@erdw.ethz.ch}}
%
%
\maketitle


\abstract{}
Astrophysical observations reveal a large diversity of radii and masses of exoplanets.  It is important to characterize the interiors of exoplanets to understand planetary diversity and further determine how unique, or not, Earth is.  Assessing interior structure is challenging because there are few data and large uncertainties.  Thus, for a given exoplanet a range of interior structure models can satisfy available data.  Typically, interior models aim to constrain the radial structure and composition of the core and mantle, and additionally ice, ocean, and gas layer if appropriate. Constraining the parameters of these layers may also inform us about interior dynamics.  However, it remains challenging to constrain interior dynamics using interior structure models because structure models are relatively insensitive to the thermal state of a planet.  Nevertheless, elucidating interior dynamics remains a key goal in exoplanetology due to its role in determining surface conditions and hence habitability.  Thus far, Earth-like habitability can be excluded for super-Earths that are in close proximity to their stars and therefore have high surface temperatures that promote local magma oceans.

\section{Introduction}

During the past two decades, numerous extrasolar worlds have been detected by ground- and space-based telescopes.  Data from the Kepler space observatory suggest that super-Earths and mini-Neptunes are among the most common planet types that occur in our stellar neighborhood \citep[e.g.,][]{petigura2013prevalence,  foreman-mackey2014,dressing2015}. The structure and composition of their interiors is largely unknown and hence even the terminology (super-Earth and mini-Neptune) may not adequately describe their interior diversity. Indeed, the large variability in super-Earth and mini-Neptune masses and bulk densities suggest a spectrum of interior structure.  \rev{For clarity and if not mentioned otherwise, we use the term super-Earth for rocky planets with small radius fractions of volatiles, and the term mini-Neptunes for planets with thick volatile layers.  Both super-Earths and mini-Neptunes are classified here as small-mass exoplanets.} 
A popular hypothesis is that different planet formation processes produce the primary building blocks that make up a planet, and the arrangement of these different components within a planet ultimately determines the planetary mass and radius.  The primary constituents that may contribute to a terrestrial planet are: (1) iron-rich core, (2) rocky mantle, (3) hydrogen-dominated gas layer accreted from the circumstellar disk, (4) heavy mean molecular weight gas layer that originates from interior outgassing (5) massive water layers.  In this chapter we focus attention on super-Earth that have small radius fractions (less than a few percent) of volatiles (gas and water); for these planets the negligible contribution of volatiles does not significantly affect the planetary mass and radius. However, we do give precursory consideration to other possible planetary interiors since we cannot necessarily confirm \emph{a priori} which interior model is most appropriate for a given exoplanet.

\begin{figure}[]
\centering
 \includegraphics[width=0.8\textwidth]{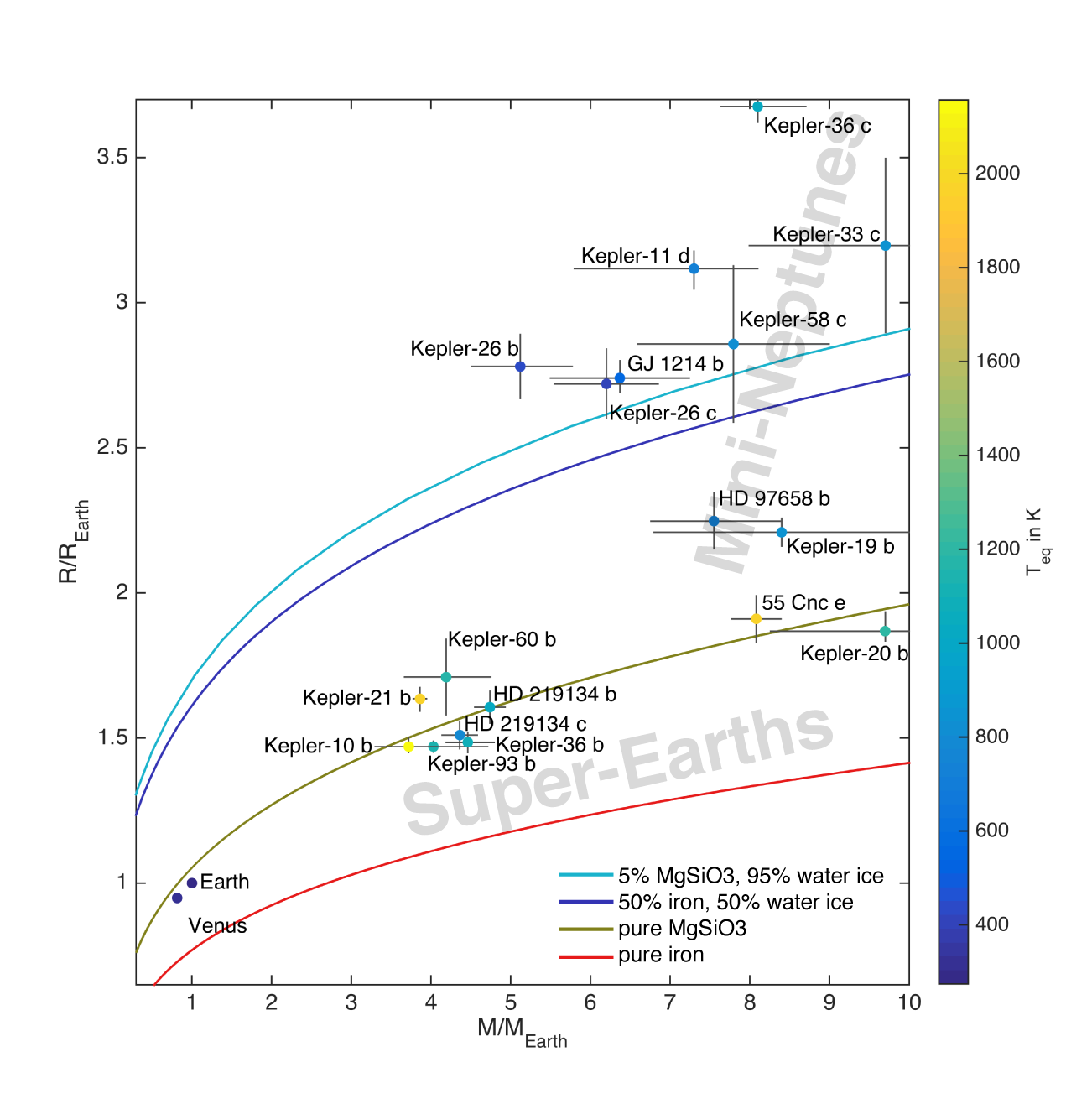}
 \caption{Mass-radius relationship of exoplanets with mass uncertainties below 20\% and the regions that broadly classify super-Earths and mini-Neptunes.  Each data point represents a planet and its color shows the planet's equilibrium temperature.  Mass-radius curves for four different compositional models are overlaid.  \label{MRplot}}
\end{figure}

The mass-radius distribution of exoplanets within the data range considered broadly appropriate for super-Earths and mini-Neptunes reveals fundamental insight into planetary diversity (Fig.~\ref{MRplot}).  The distribution of planets seems to be continuous, and super-Earths and mini-Neptunes are broadly distinguishable by their distinct increase in radii as a function of mass.  Fig.~\ref{MRplot} also shows mass-radius curves for idealized planet compositions and demonstrates that super-Earths and mini-Neptunes are consistent with rocky and volatile-rich compositions, respectively.  In essence, planets with large radii have relatively low densities which implies they have substantial amount of gas or water layers.  Super-Earth radii may be constrained to be less than 1.5--2.0 Earth radii (\RE) \citep[e.g.,][]{marcy2014,weiss2014mass,rogers2014,lopez2014understanding}, but another study suggests an upper bound of 2 Earth masses (\ME) \citep{chen2016a}.  The upper mass limit on super-Earths is based on planet formation and evolution considerations.  Super-Earths form when their mass and accretion environment prevents runaway gas accretion from the circumstellar disk, or when atmospheric loss due to stellar irradiation efficiently removes the gas layers that they previously acquired \rev{\citep{owen2016habitability,luger2015habitable}}.

Although the general relationship of mass and radius can be described by simple curves, there is considerable variability in mass and radius among super-Earths and mini-Neptunes (Fig.~\ref{MRplot}).  This variability (or scatter) is quantified by \cite{wolfgang2016probabilistic} and implies that the interior parameters that control mass and radius also exhibit variability and hence produce planetary interiors that are diverse in their composition and structure.  Important interior parameters generally include the structure and composition of core, mantle, ice, ocean and gas layers, and the internal energy of the layers.  Typically, interior models assume an iron-rich core, a silicate mantle, and H$_2$O-dominated ices and oceans, and gas (H/He or heavier elements, e.g., O, C, N).  The sensitivity of planetary mass and radius to the various structural and compositional parameters also depends on planet type, such \rev{as} whether the planet is a super-Earth, mini-Neptune, or other \citep[e.g.,][]{sotin2007mass,valencia2007detailed,howe2014mass, unterborn2016scaling}.  Furthermore, the efficiency of mixing in volatile-rich planets may influence mass-radius relationships \citep{baraffe2008, Vazan2016evolution}.

It is therefore necessary to characterize planetary interiors to understand planetary diversity.  A rigorous interior investigation needs to self-consistently account for data and model uncertainties and the likely diversity of interior parameters.  It is well-established that multiple interior models can be derived from the same mass and radius information.  For example, this ambiguity in internal structure is revealed by analyzing the parameter degeneracies using synthetic data \citep{valencia2007detailed,zeng2008computational}.  Recent work now shows that Bayesian inference analysis is a robust method for quantifying parameter degeneracy for a given (observed) exoplanet \citep{rogers2010framework, dorn2015can, dorn2017generalized}.  Inference analysis calculates confidence regions of interior parameters that relate to the probability that a planet is of a specific type.  It reveals that degeneracy is generally large, and therefore emphasizes the need to utilise extra data that informs about a planet's interior to provide strong constraints on parameters.  Thus, an objective of this chapter is to highlight astrophysical data derived from observations (other than mass and radius) that may help in this regard, such as stellar and orbital parameters, as well as spectroscopic investigations of planetary atmospheres.

The science of habitability is a young field, and it remains the subject of ongoing research to understand how we can maximize the use of interior models to inform about habitability.  However, from the perspective of interior modeling, we are primarily interested in how interior structure and dynamics facilitate the recycling of chemical components between the \rev{gas layer} and \rev{the rocky interior}.  In general, three factors seem to be key for assessing the potential for life: the availability of nutrients, energy, and liquids.  Water is expected to be the most important liquid since it is abundant on a cosmic scale, although this does not preclude the role of other liquids.  Thus it is the long-term presence of liquid water in contact with nutrient-delivering rock minerals that is regarded as a characteristic indicator of habitability. \rev{Besides the planet bulk composition, temperature and pressure conditions at the planetary surface play an important role.}  Processes that stabilize the surface temperature, such as the carbon-cycle on Earth, are key for planetary habitability \rev{\citep[e.g.,][]{kasting2010find,walker1981negative}}.  For Earth, plate tectonics is a key component of the deep carbon-cycle because carbon in the crust is subducted into the mantle and later degassing during eruption events at the surface.  Thus it is the combination of stellar irradiation and the structure and the dynamics of a planet that determine the availability of nutrients, energy, and liquids, and hence the potential for life.


In this chapter we first introduce inference analysis and discuss available data of exoplanets and how they inform us about interior parameters. We then provide a general review of interior models and give examples of interior characterization for several exoplanets.  Finally we discuss how we can link the results of structure modeling to dynamic processes and how that might inform habitability assessments.

\section{The degeneracy problem}

Interior modeling involves \rev{determining theoretical mass-radius relationships that are then compared to observed masses and radii of exoplanets in order to characterize planetary interiors} (Fig.~\ref{MRplot}) \citep[e.g.,][]{sotin2007mass, seager2007mass, fortney2013framework, dressing2014, howe2014mass}.  However, this approach alone cannot address the degeneracy problem, in which models with different interior structure and composition can have identical mass and radius.  Due to degeneracy it is challenging to understand (1) how likely an interior model actually represents a given exoplanet when frequently a large number of interior models fit the data equally well, and (2) which structural parameters are best-constrained by observations.  It is therefore necessary to address the inherent degeneracy using improved modeling techniques in order to draw meaningful conclusions about an exoplanet's interior.  One such approach is Bayesian inference analysis.
 
Inference analysis is a suitable tool to estimate interior parameters when data are sparse, the physical model is highly non-linear, or it is expected that very different models are consistent with data.  In contrast to a \emph{forward} problem that determines a result (e.g., mass-radius relationship) for a given interior description, the \emph{inference} problem consists of using measured data (e.g., mass, radius, and other observables) to constrain the parameters that characterize the interior.  The forward problem produces a unique result for a given set of interior parameters, while the inference problem provides a suite of models with different interior parameters that can explain the observations with varying degrees of success.  The inference problem requires us to explicitly quantify the known variability in parameters as \emph{a priori} information, which is determined independently from the data.  The solution of the inference problem is an \emph{a posteriori} probability distribution that reveals the sensitivity of model parameters in determining the data given the \emph{a priori} information.  Except for low-dimensional problems, this approach typically involves an extensive exploration of model parameters that requires well-designed random or pseudo-random explorations.  While stochastic sampling-based approaches for high-dimensional problems are computationally expensive, numerous global search methods exist. These include Markov chain Monte Carlo methods (McMC), nested sampling, simulated annealing, and genetic algorithms.

 A Bayesian inference for interior characterization was first performed by \citet{rogers2010framework} and various applications are described in \citet{schmitt2014planet, carter2012, weiss2015revised}.  However, the aforementioned inference is limited to few dimensions (2--3), which implies strong prior assumptions such as a planet being rocky.  Therefore, \citet{dorn2015can} devise a new method for rocky planets of general composition that permits additional data constraints and model parameters.   This method was subsequently generalized in \citet{dorn2017generalized} to include volatiles (liquid and high pressure ices, and gas layers) in models of super-Earths and mini-Neptunes.  The method employs a full probabilistic Bayesian inference analysis using McMC to simultaneously constrain structure and composition of core, mantle, ice, ocean and gas layers, as well as intrinsic luminosity of the planet.  Thus it eliminates the need for strong prior assumptions on structure and composition that were required in previous work \citep[e.g.,][]{rogers2010framework}.  Importantly, the method can utilize bulk planet constraints on refractory elements (Mg, Fe, Si, Ca, Al, Na) that are determined from stellar proxies, in addition to the usual available data of planet mass, radius, and stellar irradiation.  Furthermore, the method computes interior structure using self-consistent thermodynamics for a pure iron core, a silicate mantle, high-pressure ice, water ocean, and \rev{gas layers}.  The method is demonstrated on exoplanets for which refractory element abundances of their host stars are available \citep{dorn2017bayesian}.

\section{Observational data constraints on the interior}

\subsection {Planetary mass and radius}

Mass and radius are two fundamental parameters that can be derived from astrophysical observations (Fig.~\ref{MRplot}) and are calculated from radial velocity \rev{or transit timing variation measurements and} transit observations, respectively.  These parameters encapsulate information about the integrated interior structure and composition of an exoplanet and can be used to constrain the first-order characteristics of the interior.  Currently there are a few dozen super-Earths with measured mass and radius, but only ten or so have mass and radius uncertainties below 20\% \rev{\citep{exoplanets1}}.  Although their masses and radii can be explained by Earth-scaled interiors \citep{buchhave2016, dressing2014}, it does not preclude the existence of exoplanets with internal structures that depart from Earth's blueprint. In fact, very different interiors can result in the observed masses and radii \citep{dorn2017bayesian}.

The major constituents of a terrestrial planet (i.e., core, mantle, ice, gas) are characterized by substantial density contrasts, which enables us, to an extent, to constrain their relative layer thicknesses given planetary mass and radius (or bulk density).  The components with high density (e.g., core) influence total mass more strongly than radius, whereas the lighter components (ice, gas) strongly affect radius.  More specifically, planetary mass mainly constrains the mass fractions of core and mantle, and to a lesser extent ocean and ice layers.  By contrast, planetary radius largely dictates the thickness of volatile layers (ice, ocean, gas).  As seen from mass-radius relationships, planetary radius does not change appreciably for relatively high masses (Fig.~\ref{MRplot}).  Therefore, for high mass planets the radius can be considered as a proxy for the bulk interior structure and composition.  In this part of the domain, two planets with different mass and comparable radius may have a similar interior structure, whereas two planets with different radius and similar mass do not.

Several studies use forward models to quantify how the variability of internal parameters controls planetary mass and radius (or density) \citep[e.g.,][]{howe2014mass, unterborn2016scaling, lopez2014understanding, sotin2007mass}.  In the absence of volatile-rich layers, core size and mantle iron content dominantly affect the bulk density of the planet, whereas light elements in the core and the Mg/Si ratio of the mantle have a moderate influence \citep{unterborn2016scaling}.  If a planet harbors a gas layer, even tiny mass fractions ($<1\%$) of gas can significantly affect the radius.  Furthermore, the radius increase due to a gas layer depends on gas composition, internal energy, and the mass and size of the underlying solid interior since this controls the surface gravity \citep[e.g.,][]{howe2014mass}.  The addition of water layers can also significantly affect planetary radius given the relatively low density of water compared to rocks \citep{sotin2007mass}.  Since the presence of liquid water at a planetary surface may be a prerequisite for life, it is important to constrain the possible pressure and temperature conditions at the top of a water layer to determine the stable phase.  The surface pressure and temperature are determined by the mass and dynamics of the overlying gas layer as well as stellar irradiation.

\subsection {Stellar irradiation and age}
Since most exoplanets are detected around quiet and thus old stars (of order Gyrs), a common assumption in interior modeling is that a planet and its star are in radiative equilibrium.  The stellar irradiation that a planet receives from its host star controls the planetary surface temperature via interactions with the \rev{gas layer}.  If we consider the planet to be a perfect black body \rev{without any gas layer} then the planetary surface temperature corresponds to the equilibrium temperature of the black body, which is determined by the stellar effective temperature, stellar radius, and semi-major axis.  Most detected exoplanets orbit close to their star and hence stellar radiation is the major contributor to the available energy in the uppermost volatile layers.  This provides an opportunity to use stellar irradiation to characterize the interior structure \citep[e.g.,][]{valencia2010composition,rogers2010framework,lopez2014understanding,dorn2017generalized}.  Another energy source of lesser importance is the planet's internal heat, which originates from accretion during planet formation, core formation, gravitational compaction, radiogenic decay, and tidal heating for close-in planets with non-zero eccentricity.  The available energy in the volatile layers of a planet can significantly affect the density of gas, ocean and ice \citep{howe2014mass}.

The cooling efficiency of small-mass planets depends on the volatile content, interior dynamics, and age of the planet \citep[e.g.,][]{stevenson2003styles,fortney2006atmosphere, baraffe2008}.  The age of the planet is approximately the same age of its star, since planet formation is assumed to start soon after star formation.  Current stellar ages are inferred by comparing observations with stellar evolution models and are accurate to approximately a few orders of magnitude.  However,  considerably improved accuracy of about 10\% will soon be available through future missions such as PLATO \citep{rauer2014plato}.  Planetary age may place constraints on planetary luminosity, which is especially important for volatile-rich planets.  In principle, it is feasible to determine a range of luminosities for a given age and interior structure to compare to directly measured planet luminosities from infrared planet searches \rev{\citep[e.g.,][]{quanz2015direct}}.

\subsection {Stellar abundances}

Spectroscopic observations of stellar photospheres yield relative elemental abundances of the star \citep[e.g.,][]{hinkel2014stellar}.  Because planets and host star(s) form within the same disk, it is reasonable to expect similarities between the chemical make-up of a star and its orbiting planets.  A correlation is predicted for refractory elements (i.e., Mg, Si, Fe), since we observe that the relative abundance of these elements is comparable for the Sun, chondritic meteorites, and the terrestrial solar system planets.  The prevailing theory for chondritic meteorites is that they are the primitive building blocks of terrestrial planets and have abundances of refractory elements within 10\% of solar \citep{lodders2003solar}.  Although chondritic abundances of refractory elements can explain the bulk density of Venus, we lack independent constraints on its bulk composition, in part because we do not have rock samples from the Venusian surface.  In addition, the large relative core-size of Mercury can be explained by the removal of a large mass fraction of its silicate mantle by a giant impact \citep{benz1988}.  Therefore, the relative abundance of refractory elements is different for Mercury relative to the Sun due to the preferential removal of Mg and Si from the planet.

Numerical models of planet formation that investigate equilibrium condensation sequences argue for chemical similarities between a star and its planets.  For example, \citet{thiabaud2014stellar} show that most planets have a relative bulk refractory composition that is indistinguishable from their host star because Fe, Si, and Mg condense at \rev{a} similar temperature of around 1000 K.  Hence, most planets will replicate the refractory element ratios from the original protoplanetary disk.  Mg, Si and Fe are highly abundant elements in planet-hosting stars \citep[e.g.,][]{gilli2006abundances} and \rev{these element abundance ratios can be used as modelling constraints. Applying these constraints to interior models suggest that the mantle is the most massive layer (rather than the iron core) for the majority of rocky interiors.
Similar considerations led \citet{valencia2007detailed} to} propose that a minimum bulk ratio of Fe/Si should be applied as a constraint in interior modelling.  Similarly, \citet[][]{sotin2007mass} and \citet[][]{grasset2009study} use solar estimates of Mg/Si and Fe/Si to constrain their forward models.  Now, stellar abundances \rev{can serve as proxies for planetary bulk abundance and thus can  be used}  in Bayesian inference techniques, in addition to mass and radius, to characterize interior structure \citep{dorn2017bayesian}.  \citet{dorn2015can} quantitatively demonstrate that the generally large parameter degeneracy in interior modeling can be significantly reduced by bulk abundance constraints, which essentially introduces a strong correlation between the core size and the mantle composition. \rev{Hence these interior parameters are important for habitability assessment inasmuch as they can influence interior dynamics, cooling history, and magnetic field generation.}

\subsection {Atmospheric mass loss}

Stellar irradiation can provide atmospheric particles with sufficient energy to escape from a planet's gravitational field.  In particular, the upper part of the gas layer (atmosphere) may be subject to evaporative loss due to high irradiation.  Under the assumption that a planet has a constant reservoir of gas (steady-state) we can use evaporative mass loss considerations to put additional constraints on the structure and compositions of the interior, \rev{especially terrestrial-type planets}.  Inference analysis provides us with a suite of interior structures, and during a subsequent analysis of these models we can determine which interior structures are subject to erosion of the gas layer.  We can then further classify the models based on the stability of their gas layers \citep{dorn2017bayesian}.  This is useful because gas layers that are being lost, in much the same way as a cometary tail behaves, can be probed by transmission spectroscopy \citep{ehrenreich2015}.  This reveals that the extending atmospheres have relatively large amplitude absorptions compared to the lower atmospheric layers for a given spectral line.  These observations provide quantitative constraints on the composition and mass loss rate of the extended atmosphere \citep{bourrier2016} that we can compare with the stability analysis of gas layers from interior models \citep{dorn2017generalized}.  To demonstrate this in practice, observations do not reveal a significant hydrogen exosphere for super-Earths 55 Cnc e and HD97658b \citep{ehrenreich2015, bourrier2016}.  Hence for these planets \cite{dorn2017bayesian} exclude interior models that have thick, low-mean-molecular weight gas layers that would suffer from evaporative loss. \rev{The identification of gas layers characterized by high-mean-molecular weight can be indicative of volcanic activity. Volcanism together with crustal recycling enables greenhouse gases to cycle between the rocky interior and the gas layer and thereby can regulate surface temperature which is a key determinant in habitability assessment.}

\subsection {Tidal effects}

Orbital parameters are among the most accessible observables of exoplanets, and therefore we should exploit opportunities to constrain planetary interiors from the coupled interior-orbital dynamics of multi-planetary systems.  Exoplanets that are part of a multi-planetary system may be subject to dissipative processes such as tidal effects if eccentricities are non-negligible \citep{henning2009tidally}.  Importantly, dissipative processes can affect orbital parameters at a measurable level \citep{fabrycky2008expect}.  For the system HAT-P-13, \citet{batygin2009} demonstrate that coupling a tidal-secular orbital evolution model and an interior evolution model inform about the tidal dissipation factors of a giant planet.  Given the measured orbital parameters, the efficiency of tidal dissipation from their coupled model bounds the mass of the rocky interior of the giant planet.  Future work is needed to determine if constraints on interior structure can be derived for mini-Neptunes or super-Earths using similar tidal dissipation considerations. 


Another approach to extract information about interior structure from tidal interactions is to consider the direct influence of the star's gravitational field acting on the planet.  To prevent tidal disruption, in which gravitational forces tear a planet apart, a planet must orbit outside its so-called Roche limit.  Since the Roche limit for a given planet depends on its bulk density (in addition to other parameters), we can use observations of planets with ultra-short orbits to constrain their bulk density.  This approach was used to identify the first super-Mercury KOI-1843.03 \citep{rappaport2013}.  The authors determine a minimum density of 7 g/cm$^{3}$ for this planet.  Since the radius of KOI-1843.03  is known to within 20\%, the minimum density estimate suggests that the interior has a massive core and at most 30\% silicate mantle by mass.  Whether or not tidal considerations can be applied more broadly to characterize ultra-short period planets and planets in multi-planetary systems is subject of ongoing research.
\rev{It remains to be seen if tidal considerations can be used to distinguish between different planet types such as mini-Neptunes and super-Earths.}

\section{Interior characterization}
 
\begin{figure}[]
\centering
 \includegraphics[width = .55\textwidth]{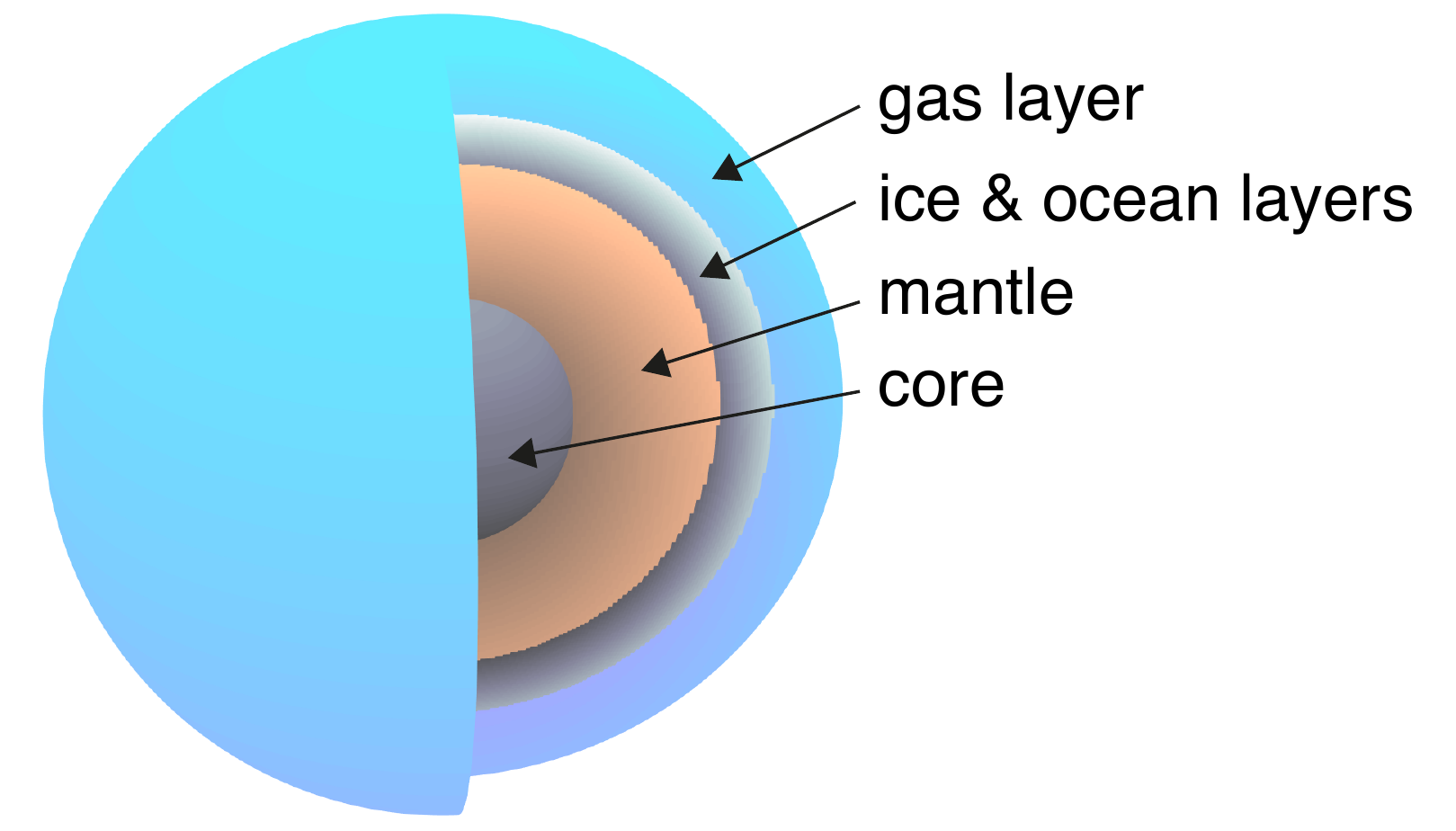}
 \caption{Schematic of a general exoplanet interior.  \label{Illustration}}
\end{figure}

\subsection{Composition}

A general exoplanet interior model consists of concentric shells that represent core, mantle, ice, ocean, and gas layers (Fig. \ref{Illustration}).  The primary objective of interior characterization is to constrain the mass and composition of each layer to obtain a first-order estimate of the hydrostatic (equilibrium) state of a planet.  There are often assumptions in models that inherently constrain the possible chemistry of the different layers; in the case of rocky planets this is typically an iron-rich core and silicate mantle.  This particular assumption is justified because rock-forming elements Mg, Si, and Fe are highly abundant in stars \citep{gilli2006abundances, hinkel2014stellar} and the interstellar medium \citep{draine2009interstellar}, which suggests that planets also form from a reservoir rich in these elements.  Iron will preferentially sink to form a core during planetary accretion and differentiation \citep[e.g.,][]{rubie2003mechanisms}.  Furthermore, we know from studying planets in the solar system that terrestrial planets are characterized by iron-rich cores and silicate mantles.  Although exotic carbon-rich compositions are proposed for planets around carbon-rich stars \citep[e.g.,][]{madhusudhan2012possible}, the material properties of carbon-rich compositions at extreme pressure conditions are the focus of ongoing research \rev{\citep[e.g.,][]{wilson2014interior, daviau2017zinc, nisr2017thermal}}.

For \rev{Earth}, seismic data gives evidence that the core is enriched in light elements, although the precise light element (or mixture of light elements) remains debated \citep{poirier1994light,badro2014seismologically}. Geochemical models justify the presence of light elements, specifically at the redox conditions at which metallic iron is stable, as many other elements are siderophile \citep[see review by][]{wood2006accretion}. With this as motivation, interior models either neglect light elements entirely in the core for simplicity, or use core compositions that were originally developed to explain Earth \citep[e.g., Fe-FeS system as in][]{sotin2007mass}.

We have learnt much about silicate mantles from study of Earth, and in particularly the character and properties of silicate mineral phases.  The Earth's mantle composition can be appropriately modeled with the ``NCFMAS'' composition model with the oxides Na$_2$O--CaO--FeO--MgO--Al$_2$O$_3$--SiO$_2$, where CaO, Al$_2$O$_3$, and Na$_2$O are minor components.  To characterize exoplanet interiors, \citet{dorn2015can} use a self-consistent thermodynamic model \citep{connolly2005computation} to calculate stable mineral assemblages for general compositions using the NCFMAS model.  The addition of minor elements enables the formation of a greater range of mineral phases but contributes little to Earth's total mass ($\sim 1\%$).  As a whole, this approach thus allows us to incorporate stellar abundance constraints in the inference analysis. 

To reduce the degrees of freedom introduced by the NCFMAS model we can instead consider a reduced set of major mineral phases: the low-pressure phases olivine ([Mg,Fe]$_2$SiO$_4$) and enstatite ([Mg,Fe]$_2$Si$_2$O$_6$), and the high-pressure phases perovskite ([Mg,Fe]SiO$_3$) and magnesiow\"{u}stite ([Mg,Fe]O).  The iron content of the silicates is defined by the Mg-number (mole fraction of Mg/(Mg+Fe)) and depends on the degree of core-mantle differentiation.  The Mg-number is unknown for most planets, except for Earth ($\sim$0.9) and Mars ($\sim$0.7).  For this reason, the models of \citet{sotin2007mass} use Earth's and Mars' value as constraints on the Mg-number of their modeled planets.  Models can be further simplified by exclusively considering MgSiO$_3$ perovskite which is the dominant silicate phase in Earth's lower mantle \citep[e.g.,][]{seager2007mass}. For guidance, assuming a pure iron core and a pure MgSiO$_3$ perovskite mantle over-predicts the mass of Earth (1.13 \ME) \citep{unterborn2016scaling}.

The major constituent of volatile layers is commonly assumed to be pure water since oxygen is more abundant in the universe than carbon and nitrogen and water condenses at higher temperatures.  The phase diagram of water has multiple phase transitions, and at high pressure liquid water is no longer stable and instead becomes a high pressure ice polymorph.  Ice VI is stable below 2 GPa and is perhaps relevant for icy super-Earths.  Ice VII and X, which are stable for pressures from $\sim 2 - 100$s GPa and temperatures from $\sim 100 - 2000$ K, are likely important for icy mini-Neptunes and ice giants \citep{french2015construction}.  In this chapter we use the term \emph{gas layer} to describe both the upper radiative \emph{atmosphere} and the underlying convection-dominated \emph{envelope}.  Gas layers can be very diverse in terms of their size, composition, and origin \citep{leconte2015anticipated}.  They can be hydrogen-dominated when they accrete from the nebular gas of the protoplanetary disk and/or be enriched with heavy elements such as C, O, and N.  Early gas enrichment is related to the disruption of accreting volatile-enriched planetesimals during planet formation \cite[e.g.,][]{fortney2013framework, venturini2016planet} as well as interior outgassing during the formation and cooling of magma oceans.  These processes are relevant during the earliest phase of planet formation and it is unclear to what extent the signature of these processes is retained in the gas layers of the low-mass planets (with ages of a few Ga) that we observe today.  The distinction between super-Earths and mini-Neptunes is often based on the radius fraction of gas layers \citep{rogers2014} which profoundly influences the total planet radius.  \rev{In addition, we expect that the gas layer for super-Earths is typically more enriched in heavy elements compared to mini-Neptunes.  This is because super-Earths have generally less mass than mini-Neptunes and therefore cannot as efficiently accrete and retain gas from the primordial disk \citep{dorn2017submitted}.}

\subsection{Structure}

Interior models consider a differentiated planet in hydrostatic and chemical equilibrium and may additionally assume that certain layers are convecting.  The interior structure of a planet can be described by mass conservation:
\begin{equation}
\frac{dm(r)}{dr} = 4 \pi r^2 \rho(r)
\end{equation}
hydrostatic equilibrium:
\begin{equation}
\frac{dP(r)}{dr} = \frac{-G m(r)\rho(r)}{r^2}
\end{equation}
and an equation of state (EoS):
\begin{equation}
\rho(r) = f(P(r),T(r))
\end{equation}
where mass $m(r)$, gravitational acceleration $g(r)$, and pressure $P(r)$, are functions of the radial distance from the planet center $r$, $G$ is the gravitational constant, and $\rho$(r) is density.  $T(r)$ is usually derived assuming that convection is the dominant heat transport mechanism such that the interior is near-adiabatic.  The adiabatic temperature gradient is:
\begin{equation}
\frac{dT(r)}{dr} = -\frac{\gamma}{K_S} \rho g T(r), 
\end{equation}
where adiabatic bulk modulus is $K_S$, and the thermodynamic Gr\"{u}neisen parameter is $\gamma$.  The relationship between the adiabatic bulk modulus $K_S$ and the isothermal bulk modulus $K_T$ is given by ${K_S}/{K_T} = 1 + \gamma \alpha T$, where $\alpha$ is the thermal expansion coefficient.  A combination of experimental and theoretical studies provide estimates of $\gamma$, $\alpha$, and $K_T$ for various materials \citep[see][]{sotin2007mass}.  An EoS describes the relationship between pressure, temperature, and density for a specific material in thermodynamic equilibrium.  A popular EoS for high pressure and temperature applications is the Birch-Murnaghan EoS, which is based on the Eulerian strain where the thermal expansion coefficient and bulk modulus are assumed to vary linearly with temperature.  Another suitable EoS for extreme planetary conditions is the Vinet EoS; For details see \citet{poirier2000introduction}.  An EoS invariably incurs error when empirical data fits at low pressure are extrapolated to high pressure.  \citet{wagner2011interior} investigate how the choice of different EoS (e.g., generalized Rydberg, Keane, and reciprocal K0) influences the interior structures that are calculated for rocky planets.  For super-Earths up to 10 \ME they find the planetary radius can differ by about 2\% as a direct consequence of the choice of the EoS.



The radial temperature profile in the gas layer is commonly simplified by considering an innermost convective layer and an outermost radiative layer.  In the radiative regime, the opacity of the layer is described by the optical depth $\tau$:
\begin{equation}
\frac{d\tau}{dm} = -\frac{\kappa}{4 \pi r^2},
\end{equation}
where $\kappa$ is opacity.  Opacities can be obtained from laboratory measurements or \emph{ab initio} calculations and opacities for different compositions are available in the literature \citep[e.g.,][for H/He or solar metallicities]{freedman2008line, jin2014planetary}.  The temperature gradient can be described by the radiative diffusion approximation:
\begin{equation}
\frac{\delta ln T(r)}{\delta ln P(r)} = \frac{3 \kappa L P(r)}{ 64 \pi \sigma G m(r) (T(r))^4}
\end{equation}
where $L$ is luminosity and $\sigma$ is the Stefan-Boltzmann constant \citep[e.g.,][]{pierrehumbert2010principles}.  The transition between the radiative and convective regime is determined by the onset of convective instabilities when the adiabatic temperature gradient is less than the radiative temperature gradient.  In order to compare structure models that incorporate a gas layer to observed radii of a given planet it is necessary to compute the transit radius of the model.  The transit radius is the distance from the planet center at which the planet is opaque to transmitted star light and can be defined by the radius where the chord optical depth is equal to one or 2/3 \citep{guillot2010radiative}.  


Due to the interaction of stellar irradiation with the gas, there are different physical and chemical processes that can significantly complicate atmospheric modeling.  Atmosphere models that account for these processes are required to interpret transmission and emission spectroscopic data (Chapter \ref{tbd} on retrieval).  However, the sparseness of available observational data does not warrant such sophisticated models for interior characterization.

\section{Examples of characterized planetary interiors}

\begin{figure}[]
\centering
 \includegraphics[width = 0.9\textwidth]{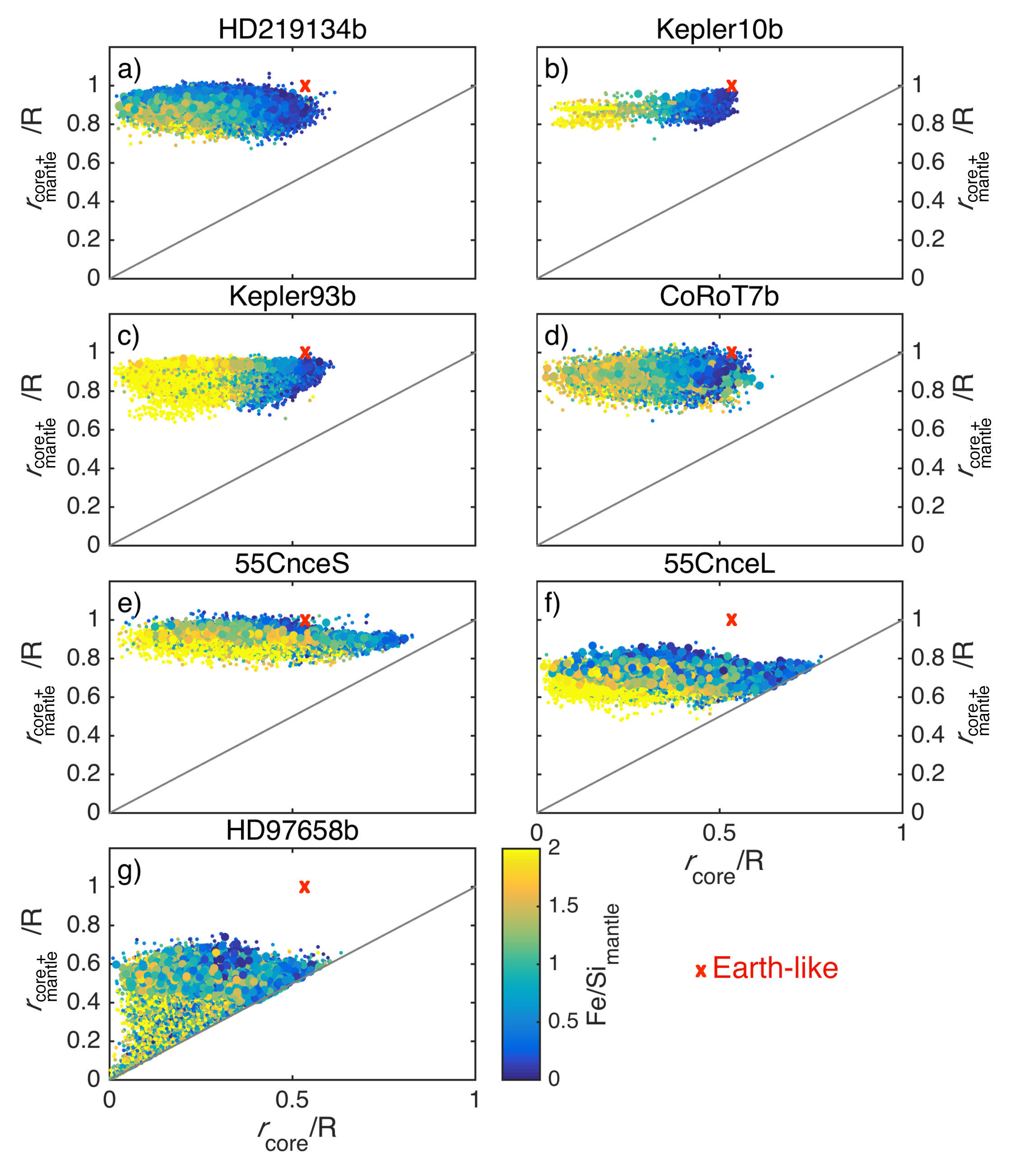}
 \caption{Sampled 2-D marginal posterior probability distributions adapted from \cite{dorn2017bayesian} for six select exoplanets (a--g).  Select interior parameters show the relative core size ($r_{\rm core}/R$), relative size of rocky interior ($r_{core+mantle}/R$), and the mantle's Fe/Si ratio, where $R$ is planet size.  The inference analysis accounts for the presence of volatile-rich layers (water and gas).  The correlation between core size and mantle composition is a consequence of stellar abundance constraints.  When abundances are well constrained, core-size and mantle iron content are anti-correlated, as illustrated by the horizontal shift in color (b--d).  When abundances are poorly constrained (a, e, f, g), the size and iron content of the mantle are anti-correlated, as revealed by vertical color shifts.  For details see \citet{dorn2017bayesian}. \label{Nr3}}
\end{figure}

\citet{dorn2017generalized} develop a generalized Bayesian inference method that allows robust comparison between the characterized interiors of different planets,.  This is because the method is applicable to a broad range of planet types that include a core, mantle, ice, ocean, and gas layer of different compositions.  We apply the method to characterize the interior of exoplanets HD~219134b, Kepler-10b, Kepler-93b, CoRoT-7b, 55~Cnc~e, and HD~97658b (Fig.~\ref{Nr3}).  The mass and radius of these planets is well-constrained, and the refractory element abundances of the stellar hosts are known \citep{dorn2017bayesian}.

Figure \ref{Nr3} illustrates that posterior distributions can vary significantly among planets depending on the specific observational data available for each planet and their associated uncertainties.  It is natural to identify a rocky planet as one which has a large radius fraction (close to one) of rock, which may include contributions from both an iron core and silicate mantle.  However, there is a strong degeneracy such that estimates of the rocky radius for a given planet can significantly depart from unity.  Unfortunately, present observational data do not provide adequate constraints on the interior of a planet to draw a definite conclusion as to whether or not a planet is rocky.  Among the characterised planets in Figure~\ref{Nr3}, Kepler-10b is the most likely to be rocky since bulk density is high and data uncertainties are relatively small.

A strength of the Bayesian framework is that it provides quantification of parameter correlations.  For example, water mass fraction is strongly correlated with the size of the rocky interior, and there is also a positive correlation between the upper range of atmospheric masses with gas metallicity. For the planets that are investigated in \citet{dorn2017bayesian}, other interior parameters are rather weakly or not correlated.
Furthermore, \citet{dorn2017generalized} and \citet{dorn2017bayesian} reveal how estimates of interior parameters depend on (1) the physical structure model, (2) prior assumptions, (3) data accuracy, and (4) data uncertainties.  For example, weakly constrained parameters (e.g., mass fraction and  composition of gas layers) are significantly affected by prior distributions, whereas strongly constrained parameters (generally thicknesses of core, mantle and gas layers) are not.  The dependence on data accuracy has been tested by using stellar abundances from different research groups with different estimation techniques \citep{hinkel2014stellar}.  \citet{dorn2017bayesian} show that stellar abundances dominantly impact the mean estimates of core size and iron content in the mantle.  An increase in the uncertainty of data generally broadens the confidence regions of interior parameters \citet{dorn2017generalized}.  Uncertainties on planetary radius mainly affect the gas layer parameters and water mass fraction.  Uncertainties on planetary mass largely influence the predicted core and mantle size, as well as water mass fraction to a lesser degree.

\section{Assessing habitability}
\subsection{Liquid surface water}

Habitability is thought to be strongly linked to the long-term presence of liquid water at the \rev{rocky} surface of a planet.  The phase diagram of water is rich in phase transitions and liquid water is confined between temperatures of 275 and 650 K and pressures between 1 kPa and 1 GPa. \rev{If a thick atmosphere (with surface pressure exceeding 100~MPa -- 1~GPa, depending on temperature) overlies a water layer, then the water will be solid rather than a liquid. Similarly,} for very massive water-layers, the depth of liquid water is limited because \rev{ice phases form} at high pressure.  For a surface temperature of 300 K, the maximum ocean depth varies from 150 to 50 km for planets of 1 and 10 \ME, respectively \citep{sotin2007mass}. The existence of a liquid water ocean in contact with a rocky interior thereby implies a maximum planetary radius.  This radius is estimated to be about 1.8 to 2.3 \RE for masses below 12 \ME \citep{alibert2014radius}, depending on the internal energy of the gas layer and the rocky interior.  \citet{noack2016water} suggest that additional oceans may exist beneath the high pressure ice phases, which would increase this maximum radius.  However, in this scenario the liquid subocean is isolated from the atmosphere which prevents study of potential biosignatures via atmospheric retrieval.

Water can also hydrate rocks. The amount of water in the rocky mantle is usually neglected for interior characterization since the mass of water in the mantle is small ($<$0.1\% by mass) and the effect on bulk density is negligible \citep{sotin2007mass}. Even for Earth, the current inventory of water in the mantle is a matter of debate and estimates span from 1 to 15 Earth oceans \citep{dai2009electrical, inoue2010water, khan2012geophysical}.  Nevertheless, water may play an important role in mantle dynamics as it can reduce the viscosity of rocks \citep{li2008water}. For Earth-like plate tectonics, partitioning of water between the surface and mantle is regulated by outgassing and subduction of weathered silicates \citep[][]{sandu2011effects, cowan2014water}. Weathering of silicates requires exposed continents.  \citet{cowan2014water} find that a tectonically active terrestrial planet of any mass can maintain exposed continents if its total water mass fraction is less than $\sim$ 0.2\%.

The challenge in identifying Earth-like planets with liquid surface water in addition to exposed continents resides in quantifying water mass fractions below 1\% \citep{cowan2014water}, which is far below observational uncertainties on planetary mass. Furthermore, the knowledge of planetary mass and radius alone are not sufficient to distinguish between thick \rev{gas layers} and oceans \citep{adams2008}. Thus, we can only exclude the presence of liquid water on top of a rocky interior when (1) the surface temperature is outside the stability field of liquid water, or (2) the bulk planet density is sufficiently low that it can only be explained with ice layers or high surface pressure and temperature due to a thick \rev{gas layer}.  To date, all small-mass exoplanets with acceptable radius and mass estimates (within 20\%) have too high surface temperature to harbor liquid oceans.

\subsection{Interior dynamics and plate tectonics}

Linking the geodynamic regime of a planet and its potential for habitability is a challenging problem in planetary science.  For Earth, plate tectonics is thought to be key for stabilizing surface temperature over geological time and thus is regarded as a potential criterion for habitability.  The classical ``mobile lid'' regime in geodynamic simulations can replicate the broad behaviour of plate tectonics, but other convective regimes exist that exhibit other surface behaviour. We briefly summarise the importance of these regimes in the context of habitability and discuss possible constraints given astrophysical data.  For more details we refer \rev{the reader} to Chapter \rev{LINK TO LENARDIC CHAPTER} \ref{tbd}.



The rocky planets in the solar system provide us with three examples of fundamentally different tectonic behaviours.  First, Mars and Mercury exhibit very little (if any) large scale surface deformation \citep{gregg15}, and therefore their mantle probably convects below a thick rigid surface known as the lithosphere.
Second, as evidenced by its cratering, the warm surface of Venus has a homogeneous age suggesting that the surface is subject to global resurfacing events \citep{strom94}.  Third, the Earth is operating in a seemingly very stable plate tectonics regime, even often called ``permanent'', in which surface material is continuously recycled into the mantle.  Understanding plate tectonics therefore requires both knowledge of the density anomalies that drive mantle convection, which are dominantly induced by lateral temperature anomalies, in addition to the mechanics of the lithosphere.

Considering the effects of temperature alone, and in particular that viscosity depends only on temperature and pressure \citep{kohlstedt15}, a stagnant (immobile) lithosphere can be produced independent of the convective vigour \citep{solomatov95}.  If we additionally assume that the lithosphere cannot sustain large stresses, geodynamic simulations can self-consistently produce a lithosphere that continuously or episodically deforms and subducts into the deep interior \citep{fowler93,tackley00b}.  The global dynamics of Venus \citep{moresi98,stein04} and Earth \citep{mallard16} can therefore be reproduced in numerical models.  However, the unknown processes that influence the yield strength of the lithosphere preclude the geodynamics community from predicting the likelihood of convection regimes.  For example, even though Venus and Earth are sister planets, they exhibit strikingly different convection regimes which defy simple explanation.  Thus, predicting the convection regime of exoplanets based solely on their measured mass and radius remains extremely challenging \citep{noack2014plate}.  While numerical models can produce a range of convection regimes, it remains difficult to predict these regimes \emph{a priori} due to the non-linear feedbacks associated with mantle dynamics.

Outgassing of the mantle can be a source of greenhouse gases that have a strong impact on the potential habitability of a planet.  Outgassing is a direct consequence of melting processes in the mantle which result in super-saturation of gaseous phases in melt when rock is adiabatically decompressed at near surface pressures.  In purely thermal convection models, the volume of rocks that experience melting and hence the quantity of gas that is released \rev{into} the \rev{overlying gas layer} is inferred from \rev{the} local pressure and temperature conditions in the mantle \rev{\citep{Xie2004,Christensen1994}}.  This is because these models do not explicitly model the transport of melt and gas.  Nevertheless, they do suggest that the mobile lid regime facilitates a continuous input of volatiles into the \rev{gas layer}.  In the so-called ``episodic regime'', lithosphere deformation occurs suddenly via catastrophic events in which the entire lithosphere sinks into the mantle.  During such resurfacing events, the production of massive amounts of melt may generate a thick secondary gas layer \citep{noack2012coupling}.  Furthermore, ingassing of volatiles in the mantle is most likely during resurfacing when large quantities of melt at the surface are directly exposed to the gas layer above.


Partial melting of mantle material has a strong impact on the cooling style of a planet because (1) melt rapidly transports heat in comparison to the slow viscous creep of solid rocks, (2) latent heat locally removes or adds heat, and (3) melt can modify the thermomechanical behaviour of the lithosphere.  For these reasons melting phenomena are often considered when investigating planetary evolution \citep[e.g.,][]{korenaga09a,kite} and the likelihood of plate tectonics.  The numerical implementation of melting and crust production, which could eventually facilitate a direct estimate of mantle outgassing, has been refined in regional and global geodynamic models during the past several decades.  Recent models have the capability to generate molten basalt and erupt it at the top of the lithosphere where it subsequently cools and solidifies \citep{xie04,keller09,moore13,gerya14,sizova15}.  \citet{lourenco16} show that crustal production has a first order effect on the dynamics of the lithosphere as it generates an additional surface load which promotes deformation.

The convection regimes that we observe in these new advanced models do not easily fit within the classical definitions that were devised based on purely thermal convection models.  Catastrophic resurfacing events now appear to be more likely than previously considered, and stagnant lids may experience significant deformation due to eruptive magmatism.  Therefore, scaling laws for the output of greenhouse gases to the \rev{gas layer} based on purely thermal convection regimes will need to be revised in light of these new results from thermochemical modeling.  Models with melting and crustal production further allow us to directly track the recycling of basalt into the mantle.  They reveal that erupted basaltic crust can be repeatedly erupted or intruded and therefore the outgassing predicted in purely thermal convection studies might be underestimated.

There is an observational bias to detect short-period exoplanets, and therefore most super-Earths that have been detected are sufficiently close to their host star that stellar irradiation strongly dictates the surface temperature.  Furthermore, these planets are most likely tidally-locked, \rev{and may be  synchronously rotating, with permanent day and night sides}.  Therefore, the dynamics of these super-Earths will presumably be significantly different from the three convection regimes (mobile, stagnant, and episodic lid) previously outlined.  This is because high surface temperature may promote the existence of a surface magma ocean on the day side and perhaps even the night side \citep[e.g.,][]{kite2016atmosphere}.  To understand these planets therefore requires us to model the dynamics of a magma ocean, which is strongly coupled to the evolution of the \rev{gas layer} \citep{elkins-tanton2008}.  This requires the advancement of numerical models to encapsulate high-melt fraction dynamics within the framework of a thermochemical convection simulation \citep[e.g.,][]{bower2017}.  Due to observational biases, detected super-Earths orbit close to their stars and therefore are subject to intense stellar irradiation. Therefore, we expect the surface conditions on these planets to be very different to Earth, which we can reasonably expect will exclude Earth-like habitability.


\section{Conclusions}

As numerous new worlds continue to be discovered, we are driven to ask how exotic are their interiors and how does Earth compare to the variability of interiors that we infer from observational data.  Our understanding of planetary diversity relies on interior characterization.
The challenge in assessing planetary interiors stems from the fact that data are limited and their uncertainties are large.  In the best case scenario, both planetary mass and radius are available for a given exoplanet.  For super-Earths there are about dozen planets with mass and radius estimates with uncertainties below 20\%.  Obtaining high quality data is of paramount importance since high precision data usually reduces parameter degeneracy.  However, there is an inherent parameter degeneracy that cannot be overcome with increasing data precision.  This is because different interior structures can have identical mass and radius.  Thus, in order to draw meaningful conclusions about a planetary interior, it is mandatory to rigorously quantify interior parameter degeneracy.  This has been successfully demonstrated by \citet{rogers2010framework, dorn2015can,dorn2017generalized} using Bayesian inference analysis.

Commonly, observed planetary mass and radius are compared to the masses and radii determined from interior models that consider a few idealized compositions.  This approach provides a useful guide for interior characterization, but it is unable to quantify the range of possible interiors that fit the data equally well.  From a few fitted interior models it is impossible to know how well they compare to the generally large number of other interior models and furthermore which interior parameters can actually be constrained by data.  There are several interior parameters that affect mass\rev{, radius, and bulk composition} and their variability should be accounted for \citep[see][for a generalized model]{dorn2017generalized}. These parameters include structure and composition of core, mantle, ice, ocean, and gas layers.  In this chapter we have provided an overview of the compositions that are typically considered for each layer and summarized general structure models.

In addition to the observations of mass, radius, and stellar irradiation, other data are provided by astrophysical observations that can be used to constrain interior models.  We have discussed the information gained by stellar abundance data, constraints from atmospheric mass loss considerations, and considered tidal forces between planet-planet and planet-star. These different data types may be used broadly (e.g., stellar abundances) or only in specific cases to characterize interiors (e.g., tidal forces). Future studies will quantify and compare the merit of different data types as well as determine the necessary precision of the data to further enhance constraints on planetary composition and structure \citep[for examples see][]{dorn2015can}.
Astrophysical observations are generally expensive, especially from space-missions, and hence thorough and careful use of available data is key for a comprehensive interior characterization.  

Interior characterization of exoplanets illustrates the challenges in identifying a specific type of interior (e.g., purely rocky interior) since degeneracy of interior parameters is generally large. For example, from mass and radius observations alone it is impossible to distinguish between the existence of oceans or thick \rev{gas layers}. So far, the best characterized planets orbit sufficiently close to their stars that surface liquid water will not be present.  In fact, there are some close-in super-Earths that have sufficiently high surface temperature that rocks are molten.  Presumably their interior dynamics are very different from the present-day Earth.


Since interior dynamics is an important contributor to habitability, we briefly review how thermochemical convection models can be used to inform about planetary surface conditions and the exchange of material between solid interior and \rev{ gas layer}.  \rev{Outgassing of an early magma ocean can lead to the formation of an enriched gas layer above a rocky planet. If plate tectonics subsequently initiates, then the long-term climate can be regulated through crustal recycling and volcanism.  This will help to maintain stable surface temperature and pressure conditions that may promote the development of life.  Furthermore, volcanism associated with plate tectonics brings both heat and nutrient-rich rock to the surface.  The desire for habitability assessment is driving the development of ever more sophisticated coupled interior-gas layer models that include realistic melt production and degassing.  Since observations cannot constrain crucial parameters that control the interior dynamics, such as rheology and thermal state, we are restricted to glean information about the dynamics from specific atmospheric signatures that are observable.  Indeed, constraints on outgassing rates and the composition of outgassed species are important goals for exoplanet atmospheric characterization and may help to constrain bulk mantle composition.  Understanding the interior structure and dynamics of rocky planets will always be a key component in assessing habitability.}

\section{Cross-References}

\begin{itemize}
\item{Atmospheric Retrieval for Exoplanet Atmospheres}
\item{Formation of Super-Earths }
\item{Tectonics and Habitability}
\end{itemize}



\bibliographystyle{spbasicHBexo}  
\bibliography{DornBib.bib} 

\end{document}